\documentclass[aoas,MSNbibl,nameyear,dvips]{arximspdf}

\doi{10.1214/10-AOAS404}
\volume{4}
\issue{4}
\pubyear{2010}
\firstpage{1634}
\lastpage{1637}

\makeatletter
\newcommand{\mc}{\mathcal}
\newcommand{\X}{\mathbf{X}}
\newcommand{\bx}{\mathbf{x}}
\makeatother

\begin{document}
\begin{frontmatter}

\title{Leo Breiman: An important intellectual and personal force in statistics,
my life and that~of~many~others\thanksref{TITL1}}
\runtitle{Leo Breiman}
\thankstext{TITL1}{Supported in part by NSF Grant DMS-09-06808.}

\begin{aug}
\author[A]{\fnms{Peter J.} \snm{Bickel}\ead[label=e1]{bickel@stat.berkeley.edu}\corref{}}
\runauthor{P. J. Bickel}

\affiliation{University of California, Berkeley}

\address[A]{Department of Statistics\\
University of California, Berkeley\\
Berkeley, California 94720-3860\\
USA\\
\printead{e1}}
\end{aug}

\received{\smonth{8} \syear{2010}}



\end{frontmatter}

I first met Leo Breiman in 1979 at the beginning of his third career,
Professor of Statistics at Berkeley. He obtained his PhD with Lo\'eve
at Berkeley in
1957. His first career was as a probabilist in the Mathematics
Department at
UCLA. After distinguished research, including the Shannon--Breiman--MacMillan
Theorem and getting tenure, he decided that his real interest was in applied
statistics, so he resigned his position at UCLA and set up as a consultant.
Before doing so he produced two classic texts, \textit{Probability}, now
reprinted as a SIAM Classic in Applied Mathematics, and \textit{Statistics}.
Both books reflected his strong opinion that intuition and rigor must be
combined. He expressed this in his probability book which he viewed as a
combination of his learning the right hand of probability, rigor, from
Lo\'eve, and the left-hand, intuition, from David Blackwell.

After a very successful career as a consultant in which he developed some
of the methods in what is now called machine learning, which became the
main focus of his research he came as a visiting professor to Berkeley in
1980 and stayed on in a permanent position till his death in 2005. As a
visiting professor he taught a course on nonparametric methods which I sat
in on. It was a question he raised in that course that led to our closer
acquaintance and subsequent collaboration. Leo had proposed goodness of
fit statistics based on the empirical process of the nearest neighbors
sphere volumes, $S_1,\ldots,S_n$ of an i.i.d. sample $\X_1,\ldots,\X_n
\sim
F$ on $\mc R^d$.

Heuristics suggested that the limiting distribution of the statistic would
be ``distribution free'' under the null if $f$ is positive and continuous.
I proposed an approach based on a variant of the ``little block,''
``big block''
technique used by \citet{rosenblatt1956central} for stationary mixing sequences.

During the year or so that we ended up spending on the paper, we found that
the heuristics were much harder to make real than I thought.
As time passed and I became testy and
grumbled to Leo, he would always comfort me with the comment that we were
plowing ``hard new ground.'' The editor of \textit{The Annals of
Statistics}, whom
I shall not name, was on a crusade to eliminate all but what he viewed as
genuinely applied papers from the journal, so he swiftly rejected the paper.
The paper was accepted by \textit{The Annals of Probability} and had
considerable follow-up
in the probability, statistics and computer science literature.

Our interests came together again in another forum: a panel to discuss, in
an unclassified fashion, problems of national security. It was then
that I was
first exposed to Random Forests (RF),
which he seemed to advocate as the cure to the
world's ills---which in many ways it was!

I very much liked the appropriateness of RF for high-dimensional data, as
exemplified in Leo's highly original approach of having each branching
of a random tree depend on a randomly selected subset of the features,
$\{ X_{ij_1},\ldots,X_{ij_p},  i=1,\ldots,n \}$ where $p$ is small
even if
$d$ is very large.

Another feature was that random trees were based on bootstrap samples, an
approach he had already examined in bagging. This had the happy feature that
for each tree there was an independent test sample corresponding to the
observations not used for that tree. These and other properties such as
importance of variables were implemented in the RF package, initiated by
Leo and completed by Adele Cutler.

It was at one of these meetings that he proposed turning unsupervised learning (Clustering)
into supervised learning (Classification), by creating a pseudo
sample from the distribution obtained by choosing features
independently according to their empirical marginal distribution. This
pseudo sample, together with the original sample to be clustered, now forms
a training sample for a two-population classification problem. For RF,
Leo then defined
a metric based on the scores of the ``true'' sample which led
to clusters. The idea is a generalization of the one he proposed for CART
where clusters correspond to leaves with a majority of ``true'' observations.
Although Leo did not arrive at this approach in the following way, it,
I believe, clarifies what's going on.

Suppose we have a sample $\X_1,\dots,\X_n$ to cluster
where $\X_i = (X_{i1},\ldots,X_{id})^{\mathrm{T}}$ has density $f$.
If we then apply RF or some other consistent classifier and permit thresholds
to vary, then for $n=\infty$, if we follow Leo's prescription,
the Bayes rule is classifying an observation $\bx$ as being from
population I, the original sample to be clustered if
\[
f(\bx) \Big/ \prod^d_{j=1} f_j(x_j) \geq c ,
\]
where the $f_j$ are the marginal densities of the $X_{1j}$, $j=1,\ldots,J$.
If $f_j \equiv1$ for all $j$, which is achievable by standardizing the
$X_{\cdot j}$ to be $\mc U(0,1)$ and $c$ is permitted to vary, this corresponds
to a method of density contours discussed by \citet{hartigan1975clustering}.
The ``product
density'' version was essentially incorporated into RF by Leo, although
he did
not develop a formal theory. Since RF reduces each coordinate to its
rank in
the sample this is all that was feasible. In any case, after standardization
only dependence determines clustering.

In the course of these meetings and a growing personal friendship, Leo
and I differed mainly in one thing: the need for regularization even
for ensemble learners. He was fascinated that for $J>1$, RF appeared to
work ``optimally'' if trees were grown to maximum purity. He held the
same view about ``boosting,'' a classification method due to
\citet{freund1996experiments}, which he championed and which he
had been the first to identify as an optimization algorithm in the
population case. Here his view was that, in the sample as well as in
population cases, this algorithm should be run as long as possible
since only improvement was possible. This was clearly false for $J=1$
for both boosting and RF since the algorithms then corresponded to the
suboptimal nearest neighbour rule. Despite evidence to the contrary
[\citet{lin2006random}], he seemed to still hope that for $d>1$,
no regularization was needed in both cases. He was certainly right that
even without regularization corresponding to pruned trees, for RF, and
stopping rules for boosting, these algorithms performed surprisingly
well.

Leo was not only a man of great originality but also strong opinions: on
regularization (as above), on the nature of modeling [\citet
{breiman2001statistical}], on the
census [\citet{breiman19941991}].
He appeared to turn against the use of mathematics in statistics but every one of
his papers contained mathematics---not general theories, but insightful
analyses with examples which corresponded to his heuristics. After a while
I became convinced that Leo loved to take extreme positions in public for
the sake of the excitement they would generate, and also for the
calling forth
of clear statements of opposing views in rebuttal to his stark statements.
I found that in
private, as expected, his views were much subtler than his public
statements.

Whatever he did, he did with gusto, proposed highly original ideas for
prediction, CART [\citet{breiman1984classification}], additive models
with Friedman [\citet{breiman1985estimating}], bagging [\citet
{breiman1996bagging}],
RF, boosting and so on.

The Berkeley Statistics Department benefited greatly from Leo's
influence as
well. He persuaded us at an early stage of the importance of developing links
with machine learning for the sake of both fields. Largely as a consequence
of his leadership, we made one stellar joint appointment with Computer Science
and Electrical Engineering which led to our present highly interdisciplinary form.

Finally, Leo drew me into the machine learning world, introduced me to NIPS,
cheered me on when I argued in my Rietz lecture about the ease of learning
when predictors concentrated on low-dimensional manifolds. He had a very
profound influence on the type of problems I have worked on during the
last 10--15 years.
Together with Yanki Ritov, we discussed semisupervised learning and realized
that this would work only if there was a coincidence between peaks and similar features
of unlabeled data and concentrations of labeled observations from
particular classes and that this advantage would become minimal as the
size of the labeled sample increased.

Shortly before his death he and I were preparing a joint proposal with Liza
Levina and Ritov for the
development of new methodology for vector multiple regression, an area
he had
already entered early with Jerry Friedman [\citet{breiman1997predicting}].
Here too, as in semisupervised learning, the theory suggested that eventually
there was no profit from using dependence between the different coordinates
of the vector to be predicted since only the conditional expectations of
the coordinates given the predictor variables mattered in the end. However,
if the number of variables to be predicted was very large, the
questions became
more interesting. Unfortunately death intervened before we could follow
these lines.

I learned a great deal from him and so did the field. His is a great personal
and scientific loss.


\printaddresses

\end{document}